\documentclass{PoS}

\title{D semi-leptonic decay form factors with HISQ charm and light quarks}

\ShortTitle{D semi-leptonic decay with HISQ action}

\author{\speaker{H.~Na}$^a$, C.~T.~H.~Davies$^b$, E.~Follana$^c$, P.~Lepage$^d$ and J.~Shigemitsu$^a$ \\ \\
   \llap{$^a$}Department of Physics, The Ohio State University, Columbus, Ohio, USA\\
   \llap{$^b$}Department of Physics and Astronomy, University of Glasgow, Glasgow, UK\\
   \llap{$^c$}Departamento de F\'{\i}sica Te\'orica, Universidad de Zaragoza, Zaragoza, Spain\\
   \llap{$^d$}LEPP, Cornell University, Ithaca, NY, USA\\ \\
        E-mail: \email{heena@mps.ohio-state.edu}}

\abstract{We present a program to study D semi-leptonic decay form factors with HISQ charm and light quarks.
In this exploratory work, we study $D_s$ to $\eta_s$, $l\nu_l$ semi-leptonic decay
on MILC coarse ($20^3\times64$) gauge configurations with 2+1 dynamical flavors.
We develop a new method to calculate $f_+(0)$ and $f_0(q^2)$ from scalar current matrix elements, $<\pi|\bar{s}c|D>$, which does not require any operator renormalization.
We also determine $f_+(q^2)$ and $f_0(q^2)$ from vector currents with a fully non-perturbative renormalization method.}

\FullConference{The XXVII International Symposium on Lattice Field Theory - LAT2009\\
		 July 26-31 2009\\
		 Peking University, Beijing, China}

\begin{document}

\section{Introduction}

Understanding the $D \rightarrow \pi$ and $D \rightarrow K$ semi-leptonic decays is an important topic in Charm physics.
Furthermore, lattice studies of $D$ semi-leptonic decays are of particular interest for several other reasons as well.
First of all, one can test lattice formulations of generic semi-leptonic decays by studying the $D$ semi-leptonic case.
For example, $|V_{ub}|$ and $|V_{cb}|$ can be determined from $B$ semi-leptonic decays, and these are very important parameters to understand CKM physics and test the Standard Model.
However, simulating $B$ mesons is, in general, more difficult than $D$ mesons.
In $B$ semi-leptonic decays, the maximum momentum transfer, $q^2_{max}$, is much larger than in $D$ semi-leptonic decays. 
In fact, one can cover the entire $q^2$ region of $D$ semi-leptonic decays on the lattice, while, for $B$ semi-leptonic decays, that would be challenging. 
In addition, the charm sector has been studied extensively  in experiments as well.
Thus, we would like to test the lattice formulation in the charm sector.
Moreover, studying D semi-leptonic decays can provide an independent determination of $|V_{cs}|$ and $|V_{cd}|$.

The $D$ semi-leptonic decay form factors, $f_+(q^2)$ and $f_0(q^2)$, can be defined as
\begin{equation}
\label{ff}
\langle \pi | V^{\mu} | D \rangle = f_+(q^2) [p_D^\mu + p_\pi^\mu - \frac{m_D^2-m_\pi^2}{q^2} q^\mu ] + f_0(q^2) \frac{m_D^2-m_\pi^2}{q^2} q^\mu,
\end{equation}
where the momentum transfer $q^\mu = p_D^\mu - p_\pi^\mu$.
Note that 
\begin{equation}
f_+(q^2=0) = f_0(q^2=0),
\label{fplus_0}
\end{equation}
because of kinematics.
This is the conventional definition of form factors, which is normally used in experiments.
For lattice calculations, it is more convenient to define the form factors as
\begin{equation}
\langle \pi | V^{\mu} | D \rangle = \sqrt{2m_D} [ v^\mu f_\| (E_\pi) + p_\bot^\mu f_\bot(E_\pi)],
\end{equation}
where $v^\mu = \frac{p_D^\mu}{m_D}$, and $p_\bot^\mu = p_\pi^\mu - (p_\pi \cdot v)v^\mu$.
This definition is particularly useful for lattice calculations in the $D$ rest-frame ($v=(1,0,0,0)$ and $p_\bot = (0,\vec{p}_\pi)$), because each form factor is determined from the temporal current matrix element or the spatial current matrix element respectively;
\begin{equation}
f_\| (E_\pi) = \frac{\langle \pi | V_0 | D \rangle}{\sqrt{2m_D}} [GeV^{1/2}], \;\;\;\;\;\;\;
f_\bot (E_\pi) = \frac{\langle \pi | V_i | D \rangle}{\sqrt{2m_D}}\frac{1}{p_\pi^i} [GeV^{-1/2}].
\label{f_perp}
\end{equation}
So, the traditional strategy is to calculate $f_\|$ and $f_\bot$ on the lattice, take the continuum and chiral limit, and then convert to $f_+$ and $f_0$ using the following relations;
\begin{eqnarray}
\label{f_zero}
f_0(q^2) &=& \frac{\sqrt{2m_D}}{m_D^2-m_\pi^2} [(m_D-m_\pi)f_\|(E_\pi)+ (E_\pi^2-m_\pi^2)f_\bot(E_\pi)], \\
\label{f_plus}
f_+(q^2) &=& \frac{1}{\sqrt{2m_D}} [f_\|(E_\pi) + (m_D - E_\pi) f_\bot (E_\pi)]. 
\end{eqnarray}
We can, then, compare the lattice calculation to experiments.
In this talk, we present a new strategy that allows us to calculate the relevant form factors from the scalar current $\langle S \rangle$ rather than the vector current $\langle V_\mu \rangle$ (Sec.~2 and 3), and the more traditional method using $\langle V_\mu \rangle$ (Sec.~4).


\section{Formalism}
We apply the HISQ action~\cite{hisq} for all valence quarks; light, strange, and charm quarks.
As we use the HISQ action, there are several noticeable advantages.
We can simulate relativistic charm quarks, since the HISQ action has small enough discretization errors~\cite{hisq}. 
In addition, we can formulate a fully non-perturbative renormalization.

Consider the following relation between the vector and scalar currents in the continuum;
\begin{equation}
q^\mu \langle V_\mu^{conti} \rangle = (m_c - m_q ) \langle S^{conti} \rangle,
\label{current}
\end{equation}
where $ \langle V_\mu^{conti} \rangle =  \langle \pi|V_\mu^{conti} |D\rangle$, $\langle S^{conti} \rangle =  \langle \pi|\bar{s}c |D\rangle$, and $q$ represents a strange or light quark.
The RHS of Eq.~\ref{current} does not require any operator matching factor (the combination is renormalization group invariant).
Thus, one can write down the same relation on the lattice with a operator matching factor $Z$ just on the LHS;
\begin{equation}
q^\mu \langle V_\mu^{lat} \rangle Z = (m_c - m_q ) \langle S^{lat} \rangle,
\label{lat.current}
\end{equation}
where $S^{lat}$ is the local scalar current.
This relation is only valid when we use the same action for all valence quarks, and the action has enough chiral symmetry.
If one uses different types of action for different quark species, then  the RHS of Eq.~\ref{current} is no longer RG invariant.
This is because the mass renormalization factors for charm and light quarks would be different, in general.

In the $D$ rest-frame, Eq.~\ref{lat.current} can be re-written
\begin{equation}
\label{z_factor}
(m_D - E_\pi ) \langle V_0 \rangle Z_t + \vec{p}_\pi \cdot \langle \vec{V} \rangle Z_s = (m_c - m_q ) \langle S \rangle,
\end{equation}
where $Z_t$ and $Z_s$ are the $Z$ factors for temporal and spatial vector current respectively.
We now omit a superscript `$lat$' on the currents.
Note that $Z_t$ and $Z_s$ can be different in general.
  Using this relation, one can extract $Z$ factors fully non-perturbatively. 
We will explain this in more detail in Sec.~4.

The scalar current matrix element can be written in terms of a single form factor $f_0$;
\begin{equation}
\langle S \rangle = \frac{m_D^2 -m_\pi^2}{m_c - m_q} f_0 (q^2).
\label{scalar}
\end{equation}
One can simply derive this relation from Eq.~\ref{ff} and \ref{current}. 
Therefore, we can obtain $f_0(q^2)$ from $\langle S \rangle$,
\begin{equation}
f_0 (q^2) = \frac{(m_c - m_q)\langle S \rangle}{m_D^2 -m_\pi^2}.
\label{fzero}
\end{equation}
The numerator of the RHS is the same RG invariant combination as in Eq.~\ref{current}, and the denominator is a mass difference of the bound-states.
Therefore, we can calculate $f_0(q^2)$ with no need for operator matching.
This is the most interesting feature due to the fact that we use the HISQ action for all valence quarks.
In addition, using this method, we can determine $f_+$ at $q^2=0$, because of Eq.~\ref{fplus_0};
\begin{equation}
f_+(0)=f_0 (0) = \frac{(m_c - m_q)\langle S \rangle}{m_D^2 -m_\pi^2}\Big|_{q^2=0}.
\end{equation}
$f_+(0)$ is a very important quantity, since we can estimate $|V_{cs}|$ and $|V_{cd}|$ by combining $f_+(0)$ and experimental inputs. 


\section{Test run for $f_0(q^2)$ and $f_+(0)$ from $\langle S \rangle$}
We have performed an exploratory analysis with MILC configurations with 2+1 dynamical flavors.
The configurations that we use are 300 coarse lattices ($20^3\times62$) with sea quark masses $u_0 am_l = 0.01$ and $u_0 am_s = 0.05$~\cite{milc_lat}.
We calculate charm and strange quark propagators using the HISQ action with quark masses $am_s=0.0546$ and $am_c=0.66$. 
For each configuration, we calculate propagators with a point source at two different time slices.
We insert four discrete spatial momenta; $\vec{p}=(0,0,0), (1,0,0), (1,1,0),$ and $(1,1,1)$.  
The non-perturbatively determined renormalization factor $\epsilon$ of the Naik term of the charm quark is $-0.21$~\cite{fds}.  

In this test run, we calculate $f_0(q^2)$ of $D_s \rightarrow \eta_s$, $l$$\nu_l$ semi-leptonic decay using Eq.~\ref{fzero}.
We use the local scalar current, and the Goldstone taste for the $D_s$ and $\eta_s$.
In the analysis, we perform Bayesian fits to the three-point function and the two-point functions of $D_s$ and $\eta_s$ simultaneously. 
We use fit model functions for two and three-point functions,
\begin{eqnarray}
C_2^{\eta}(t) &=& \sum_{n=1}^{nexp} a_n^{\eta2} (e^{-E_n^\eta t} + e^{-E_n^\eta (N_t-t)}) + \sum_{o=1}^{nexpo}  (-1)^t \bar{a}_o^{\eta2} (e^{-\bar{E}_o^\eta t} + e^{-\bar{E}_o^\eta (N_t-t)}),  \nonumber \\
C_2^{D}(t) &=& \sum_{n=1}^{nexp} a_n^{D2} (e^{-E_n^D t} + e^{-E_n^D (N_t-t)}) + \sum_{o=1}^{nexpo}  (-1)^t \bar{a}_o^{D2} (e^{-\bar{E}_o^D t} + e^{-\bar{E}_o^D (N_t-t)}),   \\
C_3(t) &=& \sum_{n=1}^{nexp} \sum_{o=1}^{nexpo} \big[ 
A_{nn} e^{-E_n^\eta t}e^{-E_n^D (T-t)} + (-1)^{T-t} A_{no} e^{-E_n^\eta t} e^{-\bar{E}_o^D(T-t)} \nonumber \\
&+&  (-1)^t A_{on} e^{-\bar{E}_o^\eta t}e^{-E_n^D (T-t)} + (-1)^{T} A_{oo} e^{-\bar{E}_n^\eta t} e^{-\bar{E}_o^D(T-t)} \big], \nonumber
\end{eqnarray} 
where the states with a bar are the negative parity states, and $T$ is the distance between the sources of the light and heavy mesons. We choose $T=19$ for this test run.
We can impose stronger constraints on the meson masses and amplitudes by taking larger fit time domains for the meson two-point functions.
Note that the correlator of $\eta_s$ at zero momentum consists of the positive parity states only.

\begin{figure}[htp]
\centering
\includegraphics[width=.37\textwidth]{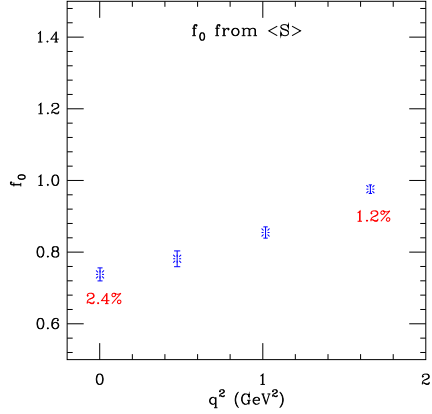}
\caption{$f_0(q^2)$ from $\langle S \rangle$.
The errors are statistical errors only. The two numbers below the data points are the relative errors.}
\label{fig:f_zero}
\end{figure}

The result is shown in Fig.~\ref{fig:f_zero}. The four data points in the figure represent 
$\vec{p}=(1,1,1)$, $(1,1,0)$, $(1,0,0)$, and $(0,0,0)$ from the left. 
We take an average over equivalent momenta to increase statistics. 
The result of $\vec{p}=(1,1,1)$ is accidentally very close to $q^2=0$, so in this particular test case, we do not need to worry about extrapolations to $q^2=0$.
We expect to achieve much smaller errors in the future, since we can improve our calculation using random-wall source techniques, more statistics, and multiple $T$ values. 

\section{Test run for the form factors from $\langle V_\mu \rangle$ with a fully non-perturbative matching}
We showed a new method that will lead to the most accurate determination of $f_+(q^2=0)$, however we are still interested in the calculation of the form factors using the traditional method to test the new method and estimate the $q^2$ dependence of $f_+$.
The traditional method was described in Sec.~1.

The vector current and pseudoscalar operator in the continuum are
\begin{equation}
V_\mu^{local} = \bar{\psi}_q(x)\gamma_\mu\psi_{q_2}(x)  \;\;\;\;\;\;\;\;\;\;\;
\mathcal{O}_{\pi}^{local} = \bar{\psi}_q(x) \gamma_5 \psi_{q_2}(x),
\end{equation}
which are local operators.
However, we cannot simply use these operators on the lattice with the Goldstone taste mesons, because we use the staggered fermion formalism.
In order to obtain non-zero matrix amplitudes, we need to consider non-local vector current or non-Goldstone pseudoscalar operators, for example
\begin{eqnarray}
V_\mu^{one-link} &=& \bar{\psi}_q(x)\gamma_\mu\psi_{q_2}(x+\hat{\mu}) 
\rightarrow  \bar{\Psi}_q(y)[\gamma_\mu \otimes I] \Psi_{q_2}(y) \nonumber \\
\mathcal{O}_{\pi}^{non-Gold} &=& \bar{\psi}_q(x) \gamma_0\gamma_5 \psi_{q_2}(x)
\rightarrow \bar{\Psi}_q(y) [\gamma_0\gamma_5 \otimes \gamma_5 \gamma_0] \Psi_{q_2}(y),
\end{eqnarray}
where $\Psi$ is the fermion field in the spin-taste basis.
In this study, we choose the spatial and temporal currents differently;
\begin{eqnarray}
\langle V_i^{one-link} \rangle &=& \langle \mathrm{Goldstone}\; \eta_s | V_i^{one-link} | \mathrm{Goldstone}\; D_s  \rangle  \nonumber \\
\langle V_0^{local} \rangle &=& \langle \mathrm{Goldstone}\; \eta_s | V_0^{local} | \mathrm{non-Goldstone}\; D_s  \rangle
\label{curr}
\end{eqnarray}
We tested the temporal matrix elements with all possible combinations of Goldstone and non-Goldstone $D_s$ and $\eta_s$, $V_0^{one-link}$, and $V_0^{local}$.
We obtained consistent results for all combinations, and the choice of Eq.~\ref{curr} gives the best results.
These combinations give the smallest statistical errors, because the taste splitting for the $D_S$ is smaller than that of the $\eta_s$.

\subsection{$f_0(q^2)$}
\begin{figure}[htp]
\centering
\includegraphics[width=.37\textwidth]{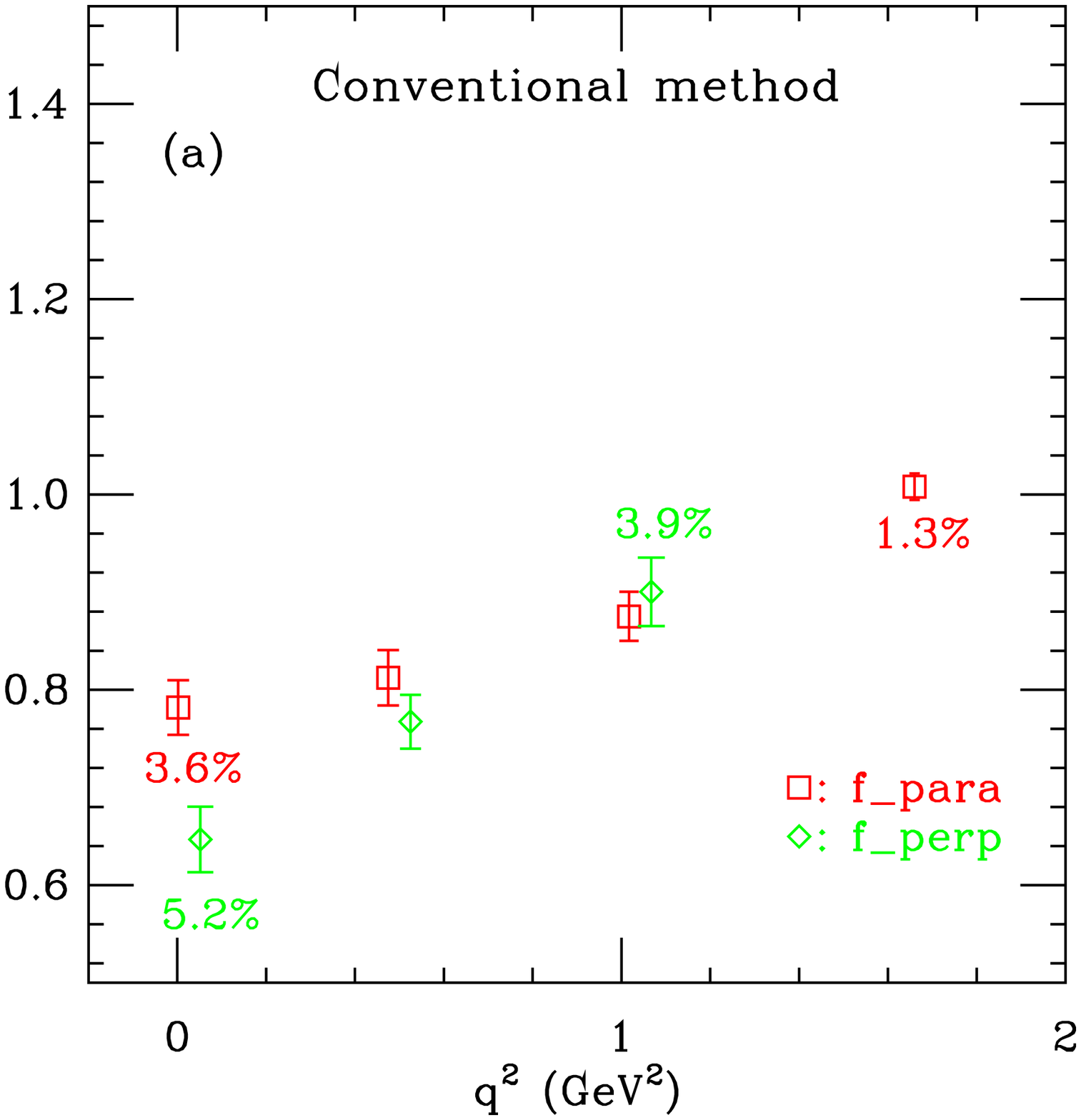}
\includegraphics[width=.395\textwidth]{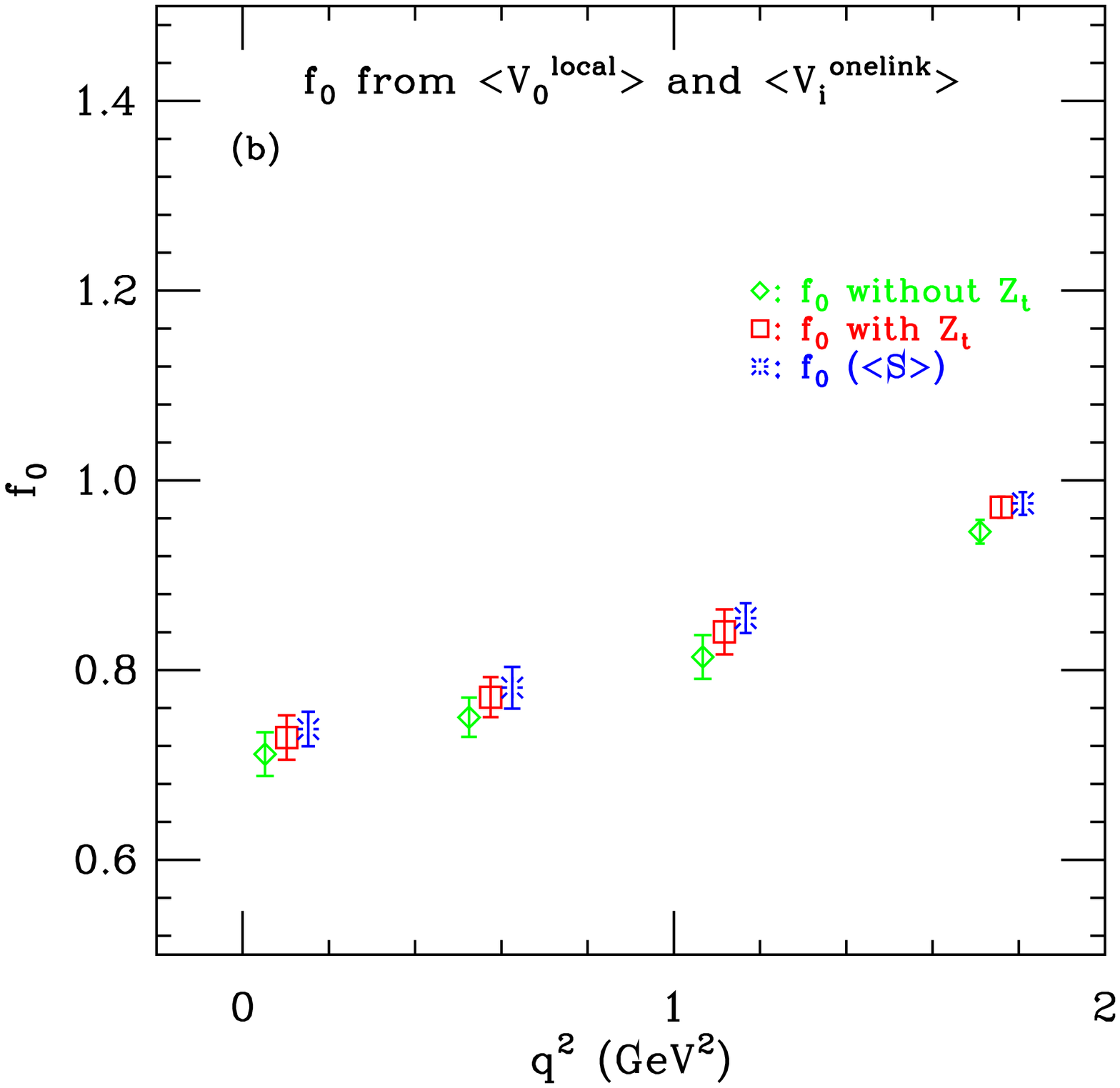}
\caption{$f_\|$ and $f_\bot$ from the traditional method are shown in (a). $f_0(q^2)$ without $Z_t$ and with $Z_t$ are shown in (b). $f_0(q^2)$ from $\langle S \rangle$ is also shown for comparison.
Form factors are calculated from the same $q^2$ values, but data points are shifted in the figures for clarity.  }
\label{fig:f_perp}
\end{figure}

In the traditional method, we need to calculate first $f_\|$ and $f_\bot$ separately as explained in Eq.~\ref{f_perp}. 
The results for $f_\|$ and $f_\bot$ before the renormalization using Eq.~\ref{curr} are shown in Fig.~\ref{fig:f_perp} (a).

After we get $f_0$ using Eq.~\ref{f_zero} (green diamonds in Fig.~\ref{fig:f_perp} (b)), we need to estimate the operator matching factors.
Taking into account that $Z_t$ and $Z_s$ can be different, we re-write Eq.~\ref{f_zero} with $Z$ factors;
\begin{equation}
f_0(q^2) = \frac{\sqrt{2m_D}}{m_D^2-m_\pi^2} [(m_D-m_\pi)Z_t f_\|(E_\pi)+ (E_\pi^2-m_\pi^2)Z_s f_\bot(E_\pi)].
\end{equation}
In fact, we do not need to consider $Z_s$ for $f_0$, since $(E_\pi^2-m_\pi^2)$ is small relative to $(m_D-m_\pi)$ and $Z_s$ is close to one.
Using Eq.~\ref{z_factor} at $\vec{p}=(0,0,0)$, we can obtain $Z_t$
\begin{equation}
Z_t = \frac{m_c-m_s}{m_D-E_\pi} \frac{\langle S \rangle|_{\vec{p}=(0,0,0)}}{\langle V_0^{local} \rangle|_{\vec{p}=(0,0,0)}} = 1.037(5).
\end{equation}
Using this $Z_t$ and $Z_s=1$, we get $f_0(q^2)$ (red squares in Fig.~\ref{fig:f_perp} (b)), which is consistent with $f_0(q^2)$ from $\langle S \rangle$.

\subsection{$f_+(q^2)$}
\begin{figure}[htp]
\centering
\includegraphics[width=.37\textwidth]{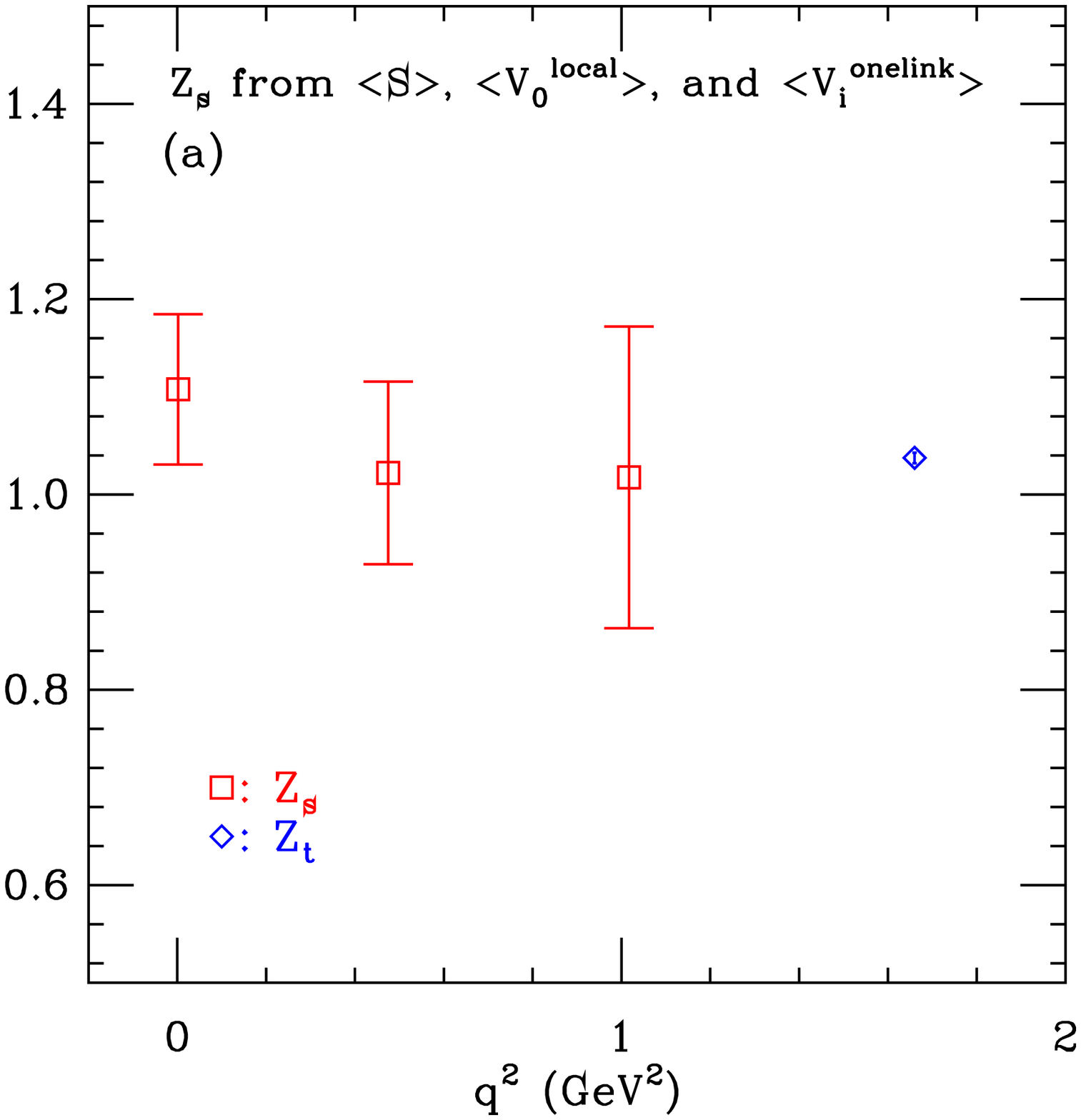}
\includegraphics[width=.395\textwidth]{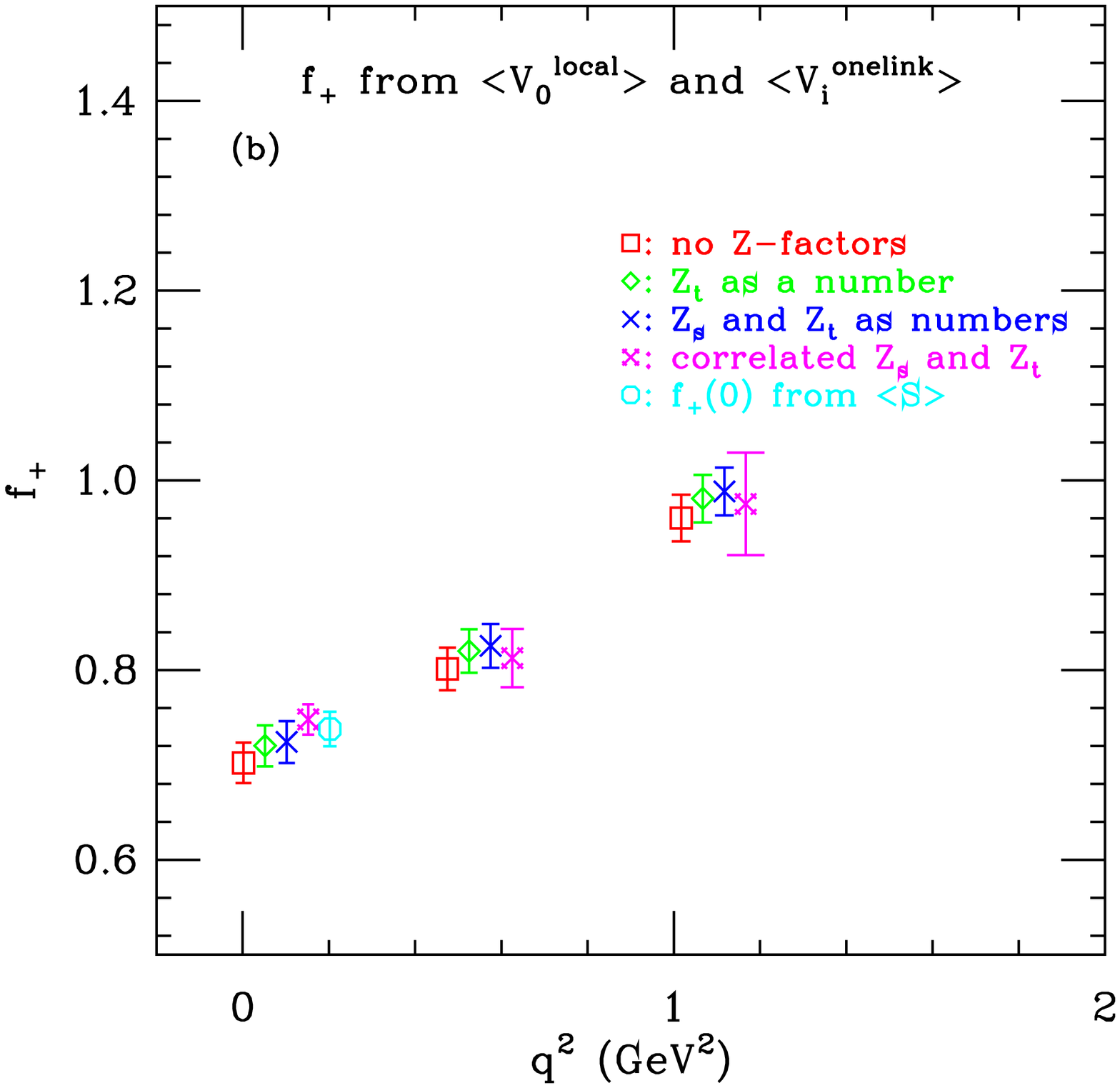}
\caption{$Z$ factors (a), and $f_+(q^2)$ from the traditional method (b).  Form factors are calculated from the same $q^2$ values, but data points are shifted in the figures for clarity. }
\label{fig:z_s}
\end{figure}

We can also re-write Eq.~\ref{f_plus} with $Z$ factors for $f_+(q^2)$,
\begin{equation}
f_+(q^2) = \frac{1}{\sqrt{2m_D}} [Z_t f_\|(E_\pi) + (m_D - E_\pi) Z_s f_\bot (E_\pi)].
\end{equation}
$Z_s$ now plays an important role for $f_+$.
When we fix $Z_t$ at $\vec{p}=(0,0,0)$, then we can obtain $Z_s$ by
\begin{equation}
Z_s = \frac{(m_c-m_s)\langle S \rangle - (m_D-E_\pi) Z_t(\vec{p}=(0,0,0)) \langle V_0^{local} \rangle}{\vec{p}_\pi \cdot \langle \vec{V}^{one-link} \rangle},
\end{equation}
since $Z$ factors are not supposed to depend on momentum.
The result is shown in Fig.~\ref{fig:z_s} (a).
As one sees, the errors of $Z_s$ are much larger than the error of $Z_t$.
This is because $Z_s$ is estimated from subtracting two positive quantities.
As a result, the error of $Z_s$ is large, even though the relative error of each term in $Z_s$ is comparable to the relative error of $Z_t$.  
The result of $f_+(q^2)$ is shown in Fig.~\ref{fig:z_s} (b).
This is a very preliminary result. We hope that we can address the best strategy to calculate $f_+(q^2)$ with correct $Z$ factors in the future.

\section{Summary and future plan}
We presented a new method to calculate form factors of $D$ semi-leptonic decays.
We showed how we calculate $f_+(q^2=0)$ and $f_0(q^2)$ with no need for operator matching in Sec.~2 and 3.
This is a general approach applicable to any semi-leptonic decays.
For instance, one can apply this method to $K \rightarrow \pi$ semi-leptonic decay.
If we have relativistic bottom quarks, we can calculate $B \rightarrow \pi$ and $B \rightarrow D^*(D)$ semi-leptonic decay form factors as well, although one will still have the problem of large $q^2_{max}$. 
The important thing is that the action should have enough chiral symmetry and all valence quarks should be calculated from the same action.
In this exploratory work, we showed that the HISQ action is quite promising for $D$ semi-leptonic decay with this new method. 
It should be possible to obtain $f_+(q^2=0)$ with significantly smaller errors than the current published theory errors of $\sim$10\%~\cite{milc_dsemi}.
It is also possible that we can replace the most accurate determination of $|V_{cd}|$ from neutrino and antineutrino scattering~\cite{pdg}.
We plan to improve our calculation by applying random-wall source techniques, adding more statistics, using smaller lattice spacings, and working with multiple $T$ values.
We will also investigate the continuum and chiral extrapolations, and extrapolations to $q^2=0$ to complete this project. 

In Sec.~4, we also presented the traditional method based on $\langle V_\mu \rangle$ with fully non-perturbative renormalization. 
We confirm that this traditional method and the new method are consistent with each other.
It gives larger errors, however it is still interesting.
Using this method, one can get the $q^2$ dependence of $f_+$.
We are still exploring other strategies for non-perturbative renormalization of $V_\mu$, which will reduce our current error on $Z_s$.

\acknowledgments{ Numerical calculations were performed on the Glenn cluster at Ohio Supercomputer Center. We are grateful to the MILC collaboration for sharing the gauge configurations. This work was supported by STFC, MICINN, NSF, and DoE.}


\begin{thebibliography}{99}
\bibitem{hisq} E.~Follana \emph{et al}., \emph{Phys. Rev. D} {\bf75} (2007) 054502.
\bibitem{milc_lat} C.~Aubin \emph{et al}., \emph{Phys. Rev. D} {\bf 70} (2004) 094505.
\bibitem{fds} E.~Follana \emph{et al}., \emph{Phys. Rev. lett.} {\bf100} (2008) 062002.
\bibitem{milc_dsemi} C.~Aubin \emph{et al}., \emph{Phys. Rev. lett.} {\bf 94} (2005) 011601.
\bibitem{pdg} C.~Amsler \emph{et al}., \emph{Phys. lett.} {\bf B667} (2008) 

\end{thebibliography}
\end{document}